\documentclass[twocolumn,preprintnumbers,amsmath,amssymb,prl,reprint,superscriptaddress]{revtex4-1}

\usepackage{graphicx}
\usepackage{color}

\newcommand{\Iexc}{I_\text{exc}}
\newcommand{\muA}{\mu\text{A}}

\newcommand{\V}{\text{V}}
\newcommand{\DAl}{{\Delta_\text{Al}}}
\newcommand{\Dstar}{{\Delta\!{}^*}}
\newcommand{\mueV}{\mu e\text{V}}

\begin{document}

\title{Transparent Semiconductor-Superconductor Interface and Induced Gap \\ in an Epitaxial Heterostructure Josephson Junction}

\date{\today}
\author{M. Kjaergaard}\thanks{These two authors contributed equally to this work}
\affiliation{Center for Quantum Devices and Station Q Copenhagen, Niels Bohr Institute, University of Copenhagen, Universitetsparken 5, 2100 Copenhagen, Denmark}
\author{H.~J.~Suominen}\thanks{These two authors contributed equally to this work}
\affiliation{Center for Quantum Devices and Station Q Copenhagen, Niels Bohr Institute, University of Copenhagen, Universitetsparken 5, 2100 Copenhagen, Denmark}
\author{M.~P.~Nowak}
\affiliation{Kavli Institute of Nanoscience, Delft University of Technology, P.O. Box 4056, 2600 GA Delft, The Netherlands}
\affiliation{QuTech, Delft University of Technology, P.O. Box 4056, 2600 GA Delft, The Netherlands}
\affiliation{AGH University of Science and Technology, Faculty of Physics and Applied Computer Science, al.~Mickiewicza 30, 30-059 Krak\'{o}w, Poland}
\author{A.~R.~Akhmerov}
\affiliation{Kavli Institute of Nanoscience, Delft University of Technology, P.O. Box 4056, 2600 GA Delft, The Netherlands}
\author{J.~Shabani}\thanks{Now at City College, City University of New York}
\affiliation{California NanoSystems Institute, University of California, Santa Barbara, CA 93106, USA}
\author{C.~J.~Palmstr\o{}m}
\affiliation{California NanoSystems Institute, University of California, Santa Barbara, CA 93106, USA}
\author{F.~Nichele}
\affiliation{Center for Quantum Devices and Station Q Copenhagen, Niels Bohr Institute, University of Copenhagen, Universitetsparken 5, 2100 Copenhagen, Denmark}
\author{C.~M.~Marcus}
\affiliation{Center for Quantum Devices and Station Q Copenhagen, Niels Bohr Institute, University of Copenhagen, Universitetsparken 5, 2100 Copenhagen, Denmark}

\begin{abstract}
Measurement of multiple Andreev reflection (MAR) in a Josephson junction made from an InAs heterostructure with epitaxial aluminum is used to quantify the highly transparent semiconductor-superconductor interface, indicating near-unity transmission. The observed temperature dependence of MAR does not follow a conventional BCS form, but instead agrees with a model in which the density of states in the quantum well acquires an effective induced gap, in our case $180 ~\mu eV$, close to that of the epitaxial superconductor. Carrier density dependence of MAR is investigated using a depletion gate, revealing the subband structure of the semiconductor quantum well, consistent with magnetotransport experiment of the bare InAs performed on the same wafer.
\end{abstract}

\maketitle

Increasing interest in the superconducting proximity effect in semiconductors arises from  recent proposals  to realize hybrid topological materials for quantum information processing \cite{Alicea:2011fe,Leijnse2011,Aasen:2015vt}. For this application, the quality of the superconductor-semiconductor interface, which controls how superconducting properties are imparted on the semiconductor, is of critical importance \cite{Blonder:1982bc,Takei:2013id,Cole:2015bl}. From another perspective, many ballistic and mesoscopic transport effects are expected in semiconductor-superconductor hybrids \cite{Beenakker:1991fv,Beenakker:1992dd} have not been investigated due to lack of an appropriate material system. Hybrid systems consisting of a superconducting metal in contact with a two-dimensional electron gas (2DEG) were widely explored in previous decades \cite{Takayanagi:1995fw,Takayanagi:1995hg,Chrestin:1997dh,Bauch:2005gh}, but material difficulties hampered progress \cite{Mur:1996er,Giazotto:2004bq,Bauch:2005gh,Amado2013}.

In semiconductor nanowires, the difficulty of creating strong uniform coupling to a superconductor was recently resolved by growing the superconductor material \emph{in situ} by molecular beam epitaxy \cite{Krogstrup:2015en}. A hard superconducting gap, measured by tunneling into the wire end, indicated an intimate coupling between materials \cite{Chang:2015kw}. More recently, \emph{in situ} growth of Al has been applied to InAs 2DEGs \cite{Javad:Xwv0duEK}. This system also exhibits a hard superconducting gap in tunnel spectroscopy \cite{Kjaergaard:yLkkp3F6}.

\begin{figure}[b]
\center
\includegraphics[width = 8.8 cm]{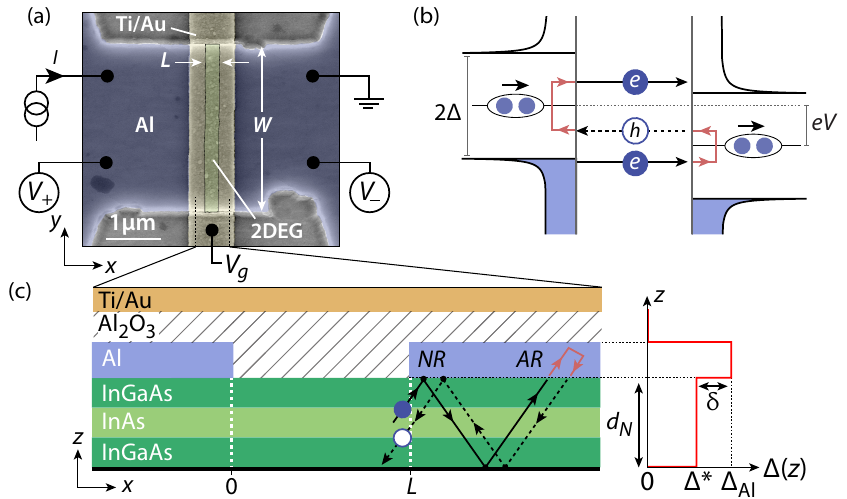}
\caption{\footnotesize{(a) False-color scanning electron micrograph of the \emph{S}-2DEG-\emph{S} device. (b) Schematic of the $2$nd order MAR process when a voltage $eV<\Delta$ is applied across the junction. (c) Cross-sectional schematic of the device in (a) (not to scale). The quantum well extends under the Al surface, allowing for processes involving multiple normal reflections (NRs) followed by an Andreev reflection (AR). Right schematic indicates variation of superconducting gap $\Delta(z)$ in the growth direction, for the case of effective quantum well thickness much less than the normal-state coherence length, $d_N \ll \xi_N$ (see text for details).}}
\label{fig1}
\end{figure}
 
In this work, we report multiple Andreev reflection (MAR) in a gateable Josephson junction formed from an InAs 2DEG/epitaxial Al heterostructure. We observe a temperature dependence of the MAR peak positions that differs from expectations for a conventional BCS-like gap, but is consistent with an induced gap in the InAs under the Al \cite{Aminov:1996uj,Takei:2013id,Cole:2015bl,Reeg:2016up}. The appearance of an induced gap, $\Dstar$, in the local density of states of the semiconductor reflects the finite time a state from the quantum well spends in the superconductor \cite{golubov1996quasiparticle}. Comparing MAR data to a quantitative model (described below), we infer an induced gap $\Dstar=180~\mueV$ and a transmission between the gapped and ungapped InAs regions in excess of $97\%$. These results are consistent with tunnel spectroscopy measurements on the same wafer \cite{Kjaergaard:yLkkp3F6}. 

The high transparency of our junction is further confirmed by the shape of MAR features appearing as peaks in {\em resistance}, rather than the more commonly observed peaks in conductance, when the voltage is tuned to $V = 2\Delta/en$. This dip-to-peak transition is a longstanding prediction for highly transparent junctions \cite{Averin:1995do}, also confirmed by our quantitative modeling. To our knowledge, this inversion has not be reported in the experimental literature, even for junctions considered highly transparent (see, for instance, Ref.~\cite{Doh:2005cf,Xiang:2006bg,Li:2016cb}). We discuss the dip-to-peak transition further in the Supplemental Material \cite{supplement}.

Modeling also reveals the existence of two distinct families of MAR resonances at zero top-gate voltage, which we associate with two occupied subbands in the 2DEG. By energizing a top gate on the exposed 2DEG, the resonant features change, becoming consistent with single-subband occupancy. The gate-dependent change from two to one subband is consistent with magnetotransport measurements on a Hall bar with the Al removed, fabricated on the same wafer. 

The hybrid heterostructure was grown by molecular beam epitaxy, and consists (from top to bottom) of 10~nm Al, 10~nm In$_{0.81}$Ga$_{0.19}$As, 7~nm InAs (quantum well), 4 nm In$_{0.81}$Ga$_{0.19}$, and an InAlAs buffer on an InP wafer. We emphasize that the Al layer is grown \emph{in situ} as part of the heterostructure \cite{Javad:Xwv0duEK}. Density $n=1.26\times10^{12}~\text{cm}^{-2}$ and mobility $\mu = 15600~\text{cm}^2/\text{Vs}$, measured in a top-gated Hall bar geometry with the Al removed, yield a mean free path $l_e\sim 290~\text{nm}$ at top-gate voltage $V_g= -2.5~\text{V}$. As demonstrated below, at $V_g= 0$ the quantum well has two subbands occupied (see also Supplementary Material \cite{supplement}).

The wafer is patterned into mesa structures using a standard III-V wet etch. The Al is then patterned using a selective Al etch (Transene D). Next, an unpatterned $40$~nm aluminum oxide layer is deposited using atomic layer deposition. Finally, a Ti/Au gate is deposited, patterned to cover the exposed 2DEG. Figure \ref{fig1}a shows a false-color scanning electron micrograph of the final device, and Fig.~\ref{fig1}c shows a schematic cross-section through the junction. The exposed 2DEG region has a length $L\simeq 250~\text{nm}$ and a width $W= 3~\mu\text{m}$. The superconducting gap of the 10 nm thick Al layer is inferred from the critical temperature ($T_c = 1.56~\text{K}$, independently measured in four-terminal measurement) via $\DAl = 1.76\,k_BT_c = 237~\mueV$. We note that the gap of the Al layer is larger than bulk Al \cite{Merservey1969}, with a $T_c$ consistent with previously reported values \cite{Meservey:1971tg,Court:2008gu}.

All measurements were performed in a dilution refrigerator with base temperature  $T \sim 30~\text{mK}$ using standard DC and lockin techniques, with current excitation in the range 2.5~nA to 5~nA. 

\begin{figure}[t]
\center
\includegraphics[width = 8.8 cm]{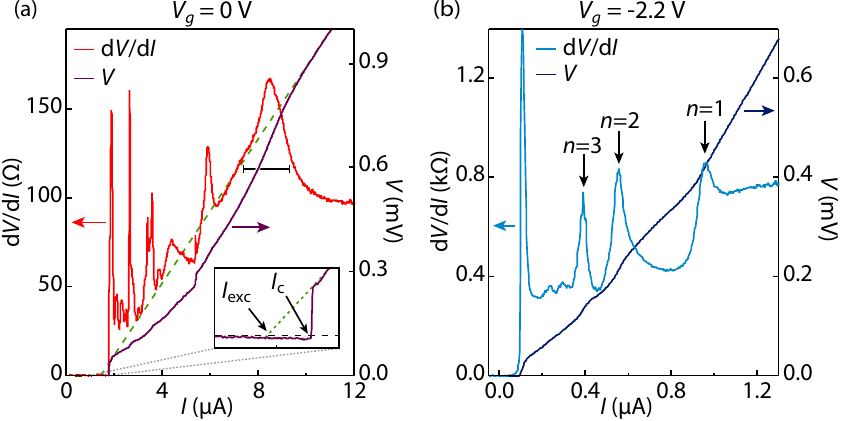} 
\caption{\footnotesize{Differential resistance (left axis) and voltage (right axis) at two different gate voltages. In (a), the dashed green line shows linear fit at $eV(I) \gg \DAl$, used to extract the excess current, $\Iexc$, as the intercept with the $V=0~$mV (as shown in the inset). Abscissa on the inset ranges from $1~\mu\text{A}$ to $2~\mu\text{A}$. $I_\text{c}$ is the current at which the system switches to a resistive state. The peaks highlighted in (b) correspond to multiple Andreev reflections of order $n$.}}
\label{fig2}
\end{figure}

The theoretical approach to this system begins with the Octavio-Blonder-Tinkham-Klapwijk (OBTK) model for multiple Andreev reflections \cite{Octavio:1983ex}. As originally formulated, this model assumes a well-defined voltage is dropped across the normal region (green rectangle in Fig.~\ref{fig1}a), leading to the MAR process sketched in Fig.~\ref{fig2}b. For a planar junction where the 2DEG extends {\em under} the Al (Fig.~\ref{fig1}c), the voltage can also drop along the horizontal Al-2DEG interface. In the case of imperfect Al-2DEG transparency, this leads to smearing of the resonances arising from MAR \cite{Kleinsasser:1990ff,Nitta:1992ea}. The OBTK model was later extended to account for the planar geometry \cite{Aminov:1996uj}, denoted \emph{SNcNS}, where $c$ is the semiconducting region in which the superconducting top layer has been removed. The \emph{SN} electrodes, consisting of 2DEG with Al on top, are assumed to be disordered and in equilibrium, while the exposed 2DEG region of length $L$ is assumed ballistic. The model yields a renormalized density of states in the 2DEG, with an induced gap, $\Delta^* < \DAl$ determined by the quality of the interface between the 2DEG and the Al~\cite{Aminov:1996uj}. 

Figure \ref{fig2} shows differential resistance (left) and DC voltage (right), as a function of applied DC current, for two gate voltages. The inset in Fig.~\ref{fig2}a show a zoom-in indicating the excess current and critical current for $V_g = 0~\text{V}$. The critical current is $I_c = 1.77~\muA$ yielding an $I_cR_n$ product of $165~\mueV$, about 70\% of the gap of the Al film, and a critical current density $J_c = 0.59~\mu\text{A}/\mu\text{m}$. The excess current, reflecting enhanced current through the junction due to Andreev reflection, is defined as the $V = 0$ intercept of a linear fit to $V(I)$ taken at $V\gg \DAl/e$ (green dashed line in Fig.~\ref{fig2}a). The measured excess current, $\Iexc = 1.44~\muA$, corresponds to $\Iexc R_n = 140~\mueV$ \footnote{The excess current is related to the gap via $\Iexc = \alpha \Delta/eR_n$, where $\alpha = 8/3$ in the ballistic, fully transparent case \cite{Blonder:1982bc}, and $\alpha = (\pi^2/4-1)$ in the diffusive case \cite{Artemenko:1979kb}.}. The differential resistance (red curve in Fig.~\ref{fig2}a) shows a series of peaks as the current is increased. The peak/dip structure is a manifestation of the MAR processes and is expected to follow the series $eV = 2\Delta/n$, with $n = 1,2,3,...$ corresponding to the number of Andreev reflections.

However, the broad resistance peak, highlighted with black horizontal bar in Fig.~\ref{fig2}a, occurs at energies larger than $2\Delta$ but follows the temperature dependence of $I_c$ and disappears at $T_c$ (see Supplemental Material \cite{supplement}), indicating that the feature has a superconductive origin. Such anomalous resistance features are believed to be associated with the planar Josephson junction geometry \cite{Nguyen:1992dr,Nguyen:lPrYu5nv}, where quasiparticles in the 2DEG can undergo several scattering events at the Al interface before ultimately undergoing Andreev reflection and traversing the same path back. An example of such a process is sketched in Fig.~\ref{fig1}c. On a length scale smaller than the normal-state coherence length $\xi_N = \hbar v_F/k_B T$, this process will appear as Andreev reflection from an \emph{effective} boundary, indicated by the white dashed line in Fig.~\ref{fig1}c. The finite-bias properties of such systems cannot be adequately described by either the \emph{SNcNS} or OBTK models, and the simple picture in Fig.~\ref{fig1}b breaks down. With the contacts out of equilibrium, the position of the peaks in Fig.~\ref{fig2}a cannot be directly related to the superconducting gap. However, by increasing the resistance in the exposed region relative to the horizontal interface, the peaks at finite-bias follow a regular series and can be used to extract a value for the induced gap.

In Fig.~\ref{fig2}b, the gate covering the exposed 2DEG region is energized to $V_g = -2.2~\text{V}$, substantially depleting the junction, leading to a normal state resistance $R_n = 740~\Omega$. At this gate voltage, the broad resistance peak at energy $eV > 2\DAl$ is absent, and the DC voltages of the first three peaks (indicated with vertical black arrows) are positioned proportional to $1/n$, indicating that the voltage drop now occurs predominantly in the uncovered 2DEG region. At this gate voltage $I_cR_n$ is reduced from the $V_g = 0$ value (full $I_cR_n$ versus $V_g$ given in Supplementary Material \cite{supplement}). As we show below, the $IV$ curves in Fig.~\ref{fig2} are consistent with near unity transmission through the junction. In such transparent junctions, the peaks due to MAR resonances appear in the differential resistance, as opposed to peaks in the differential conductance \cite{supplement}. The vertical arrows in Fig.~\ref{fig2}b therefore point to peaks in resistance, not in conductance, to indicate multiples of the gap, arising from the relation $V = 2\Delta/en$

\begin{figure}[!ht]
\center
\includegraphics[width = 8.8 cm]{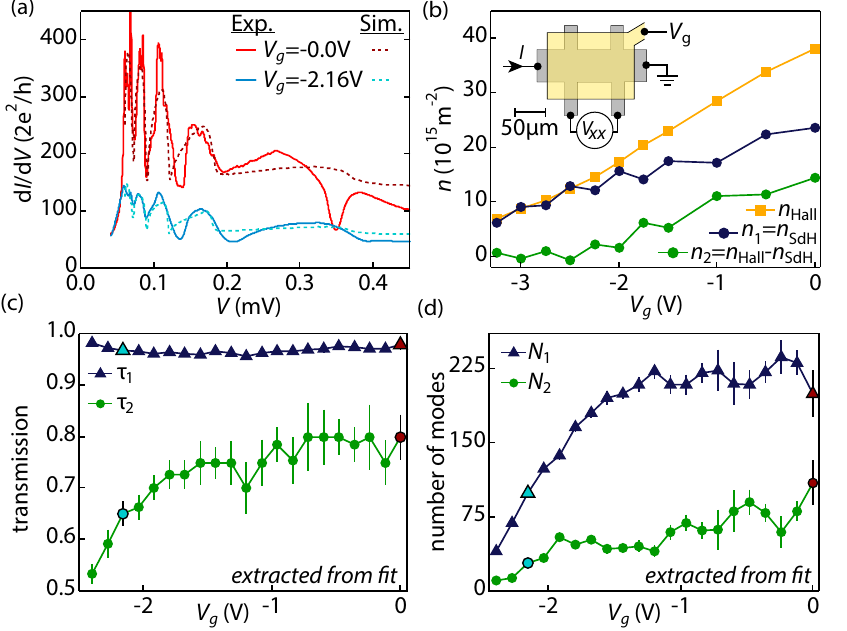}
\caption{\footnotesize{(a) Conductance as a function of bias voltage at two different gate voltages exhibiting resonances due to MAR. (b) Density in the 2DEG extracted from Hall slope and power spectrum of Shubnikov-de Haas oscillations versus gate voltage (See text for details). (c),(d), Transparency, $\tau_i$, and number of modes, $N_i$, in subband $i$, as a function of top gate voltage $V_g$, extracted from the MAR data in (a). The red and teal points correspond to the fitting values used for the dashed curves in panel (a).}}
\label{fig3}
\end{figure}

To extract the value of $\Dstar$, we plot the conductance from Fig.~\ref{fig2} against the DC voltage drop, as shown in Fig.~\ref{fig3}a. The theoretical MAR resonances in Fig.~\ref{fig3}a are simulated using a generalized scattering matrix approach developed for \emph{SNS} junctions \cite{Averin:1995do,Bratus:1995bc}. Within the model of an induced gap \cite{Aminov:1996uj} the \emph{SNcNS} system is interpreted as an effective \emph{S}$^*\!$\emph{N}\emph{S}$^*$-junction, where $S^*$ is the superconducting quantum well with a gap $\Dstar$ and a critical temperature identical to that of the parent superconductor. Simulations are performed by calculating the conductance $G^{(\tau)}(V)$ of a single mode with transmission $\tau$, from the DC component of the current $I^{(\tau)}(V,t) = \sum_k I_k^{(\tau)} \exp(2ikeVt/\hbar)$. The time-independent Fourier component of $I_k$ is calculated from the wave functions of the quasiparticles accelerated by the voltage $V$ across the junction. In the case of a ballistic junction ($L < l_e$), the back-scattering effectively only occurs at the boundary between $S^\star$ and $N$ (dashed white line in Fig.~\ref{fig1}c). The total current through the junction is the sum of currents carried by $N$ modes in $M$ subbands. The resulting conductance through the multimode junction is given by $G(V) = \sum_i^M N_iG^{(\tau_i)}(V)$ where $N_i$ is the number of modes in the $i$'th subband, and $\tau_i$ is the transmission of the modes in the $i$'th subband. 
 
A nonlinear least-squares procedure is used to fit simulated $G(V)$ curves to the data in Fig.~\ref{fig3}b, where $\tau_i, \Delta^\star$ and $N$ are fitting parameters and $M$ is predefined (see also \cite{Allen2015}). The minimal number of subbands needed to capture the essential features of the data was found to be $M=2$. For $M>2$ the optimal fit did not populate the $i>2$ subbands (i.e. $N_i \sim 0$ for $i>2$), indicating that the data is well described by two subbands (in Supplementary Material we present simulations using $M=1$ and $M=3$ \cite{supplement}). The result of fitting to the MAR features at two $V_g$ values are shown as dashed curves in Fig.~\ref{fig3}a. At $V_g=0$, the induced gap was $\Dstar = 182~\mu\text{eV}$ with $N_1 = 199$, $N_2 = 109$, $\tau_1 = 0.98$, and $\tau_2 = 0.8$. When the gate is energized to $V_g = -2.2~\text{V}$ the fitting values are $\Dstar = 180~\mu\text{eV}$, with $N_1 = 100$, $N_2 = 29$, $\tau_1 = 0.97$, and $\tau_2 = 0.65$. The gate-voltage dependence of the fitting parameters $\tau_i$ and $N_i$ are shown in Figs.~\ref{fig3}c and \ref{fig3}d. The gap $\Dstar$ extracted from the fitting routine is identical to the one measured in a tunneling experiment on the same wafer \cite{Kjaergaard:yLkkp3F6}.

The presence of two transmission species in the optimal fit is attributed to the 2DEG having two occupied subbands. The carrier density in the 2DEG, denoted $n_\text{Hall}$, is measured in a Hall bar geometry via the Hall slope (shown in Fig.~\ref{fig3}b). The density from the Hall slope is compared to the density extracted from the periodicity of the SdH oscillations in an out-of-plane magnetic field. The data in Fig.~\ref{fig3}b show the density change in the 2DEG as the top gate is energized. The power spectrum of $\rho_{xx}(1/B)$ exhibit a two peak structure, indicating two subbands with different densities in the quantum well at $V_g = 0~\text{V}$ \cite{Averkiev:2001iq}. The density corresponding to the major peak is denoted $n_1$, and the difference $n_\text{Hall} - n_1$ is denoted $n_2$. The density in the two subbands changes as the topgate is energized, as shown in Fig.~\ref{fig3}b, similar to $N_1$ and $N_2$ extracted from fitting to the MAR features. In particular, the $N_2$ species becomes depopulated at a gate voltage similar to the depletion of the second subband in the Hall bar (Fig.~\ref{fig3}b). The decrease of transmission of the $i=2$ species in Fig.~\ref{fig3}c could be due to a breakdown of the ballistic assumption as the second subband is depleted.

Within the 1D Blonder-Tinkham-Klapwijk (BTK) formalism for an \emph{SN} interface, the  transparency is often parametrized using the dimensionless quantity $Z$, related to the transmission via $\tau^{-1} = (1+Z^2)$ \cite{Blonder:1982bc}. For the first subband we extract an average transmission $\bar\tau_1 \gtrsim 0.97$, corresponding to a $Z$-parameter of $Z_1 \lesssim 0.18$, indicating a near pristine \emph{effective} interface for this mode, between the uncovered and Al-covered regions of the 2DEG in this system.

The distinction between a BCS-like gap, $\DAl$, and an induced gap, $\Dstar$, is revealed through the temperature dependence of the superconducting properties. In the case where the effective thickness of the quantum well is much less than the normal-state coherence length, $d_N \ll \xi_N$, any position-dependence of the gap magnitude in the growth direction in the 2DEG can be neglected, and the temperature dependence of the induced gap depends on $\DAl$ according to \cite{Volkov:1995ip,Aminov:1996uj,Chrestin:1997dh}
\begin{equation}
\Dstar(T) = \frac{\DAl(T)}{1+\gamma_B\sqrt{\Delta_\text{Al}^2(T) - {\Delta^*}^2(T)}/\pi k_B T_c},
\label{eq:inducedgapTdep}
\end{equation}
where $\DAl(T)$ is determined self-consistently from BCS theory. The dimensionless parameter $\gamma_B$ is a measure of the horizontal \emph{SN} interface transparency, where $\gamma_B = 0$ corresponds to a perfectly transparent interface \cite{Schapers:2001dv}. The parameter $\gamma_B$ represents the discontinuity in the superconducting pair-potential and gives rise to the difference between the gap in aluminum, $\DAl$, and the induced gap, $\Dstar$, in the 2DEG, denoted $\delta$ in Fig.~\ref{fig1}c. 
For the present case we find $\gamma_B=0.87$, using $\Dstar = 180~\mu\text{eV}$ and $\DAl =237~\mu\text{eV}$, consistent with a high quality interface between the semiconductor and Al.

\begin{figure}[!ht]
\center
\includegraphics[width = 8.8 cm]{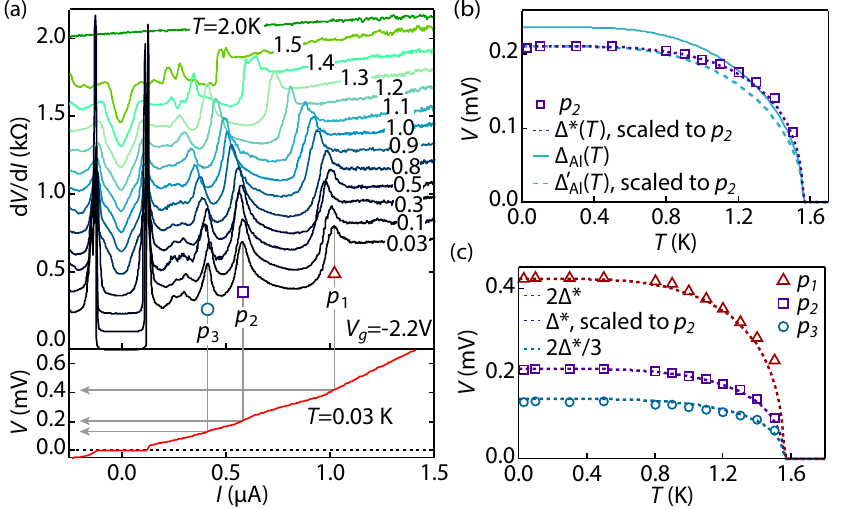}
\caption{\footnotesize{(a) Temperature dependence of the MAR features at $V_g = -2.2~\V$. Traces successively offset by $120~\Omega$. Bottom panel shows sample $IV$ curve at $T = 0.03~\text{K}$, used to extract voltage drop at each peak. (b) Temperature dependence of the peak labeled $p_2$. Dashed purple line is Eq.\eqref{eq:inducedgapTdep} scaled to match $p_2$ at base temperature. Solid teal line is temperature dependence of a BCS superconducting gap, and dashed teal line is a rescaling of $\DAl (T)$, to match $p_2$ at base temperature. (c) Temperature dependence of first, second, and third peak positions, with multiples of $\Dstar(T)$ from (b).}}
\label{fig4}
\end{figure}

To elucidate the nature of the induced superconducting gap, we study the temperature dependence of the differential resistance at $V_g = -2.2~\text{V}$, shown in Fig.~\ref{fig4}a. The position of the second MAR related peak (denoted $p_2$) is tracked in Fig.~\ref{fig4}b as the temperature is increased.  The curves in Fig.~\ref{fig4}b show the solution of Eq.~\eqref{eq:inducedgapTdep} (purple), temperature dependence of a BCS gap, $\DAl(T)$, (teal), and a BCS-like gap, $\Delta_{\text{Al}}'(T)$, (teal, dashed), where the gap value has been rescaled to coincide with the data at $T=30~\text{mK}$. The inadequacy of the temperature dependence of a BCS-like gap (both unscaled and rescaled) to account for the temperature dependence of the peaks is contrasted by the good correspondence between Eq.~\eqref{eq:inducedgapTdep} and our data. The temperature dependence of the first and third peak positions, $p_1$ and $p_3$, are shown in Fig.~\ref{fig4}c. The curves identified with $p_1$ and $p_3$ are found by multiplying $\Dstar(T)$ by a factor of 2 and 2/3, respectively, corresponding to $n=1$ and $n=3$ in the $2/n$ MAR series. We note that in the case of a highly transmissive junction the position of the maximum related to the $n=2$ MAR resonance is not located exactly at $\Delta/e$ \cite{supplement}. The excellent agreement also with $n=1$ and $n=3$ resonances indicate that the superconducting properties of the junction are well described within the induced gap model.

In conclusion, we have measured MAR resonances in a Josephson junction in a InAs 2DEG heterostructure, where aluminum is epitaxially matched to the 2DEG. By fitting the conductance of the MAR features, we extract a transmission close to unity through an effective \emph{S}$^*\!$\emph{N}\emph{S}$^*$-junction, where \emph{S}$^*$ represents the InAs quantum well covered by the Al. The temperature dependence of the MAR resonances is well-described by the theory of an effective induced gap, and we find $\Dstar = 180~\mueV$ in the Al covered 2DEG region.

\emph{Acknowledgements:} The authors acknowledge enlightening discussions with K. Flensberg. Research support by Microsoft Research and the Danish National Research Foundation. C.M.M. acknowledges support from the Villum Foundation. F.N. acknowledges support from a Marie Curie Fellowship (No.~659653). M.P.N. acknowledges support from ERC Synergy Grant. A.A. is supported by an ERC Starting Grant, the Foundation for Fundamental Research on Matter (FOM) and the Netherlands Organization for Scientific Research (NWO/OCW) as part of the Frontiers of Nanoscience program.

\bibliographystyle{apsrev4-1}
\bibliography{inducedgap_mar}

\clearpage
\onecolumngrid
\begin{center}
\textbf{\large Supplementary material for ``Transparent Semiconductor-Superconductor Interface and Induced Gap in an Epitaxial Heterostructure Josephson Junction''}
\end{center}
\appendix

\setcounter{equation}{0}
\setcounter{figure}{0}
\setcounter{table}{0}
\makeatletter
\renewcommand{\theequation}{S\arabic{equation}}
\renewcommand{\thefigure}{S\arabic{figure}}

\subsection*{Mobility peak and Shubnikov-de Haas oscillations}
\begin{figure}[!ht]
\center
\includegraphics{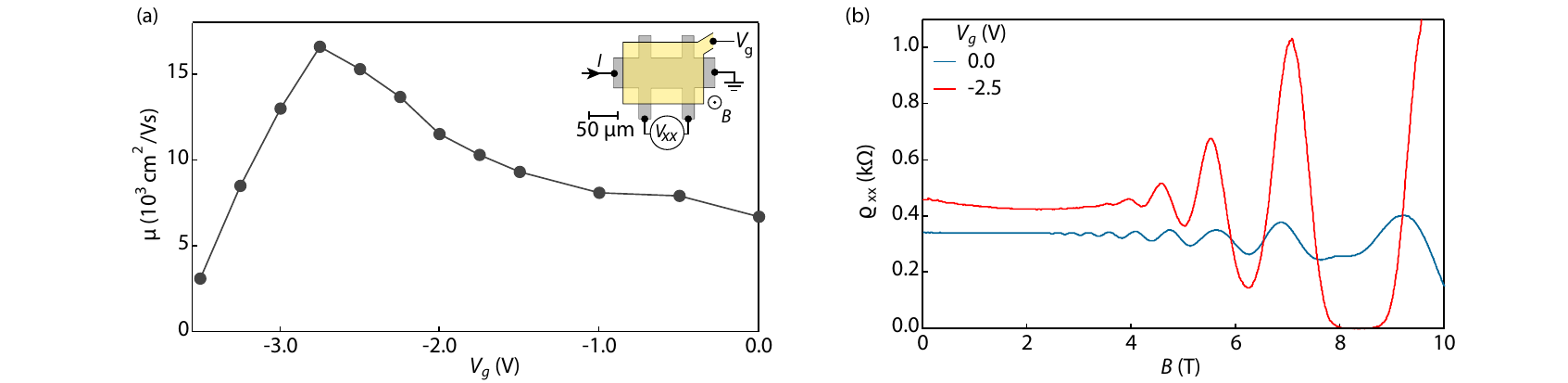}
\caption{\footnotesize{(a) Mobility of the 2DEG as a function of top gate voltage measured in a Hall bar. (b) Shubnikov-de Haas oscillations at two values of top gate voltage in a Hall bar.}}
\label{fig:mobility_and_sdh}
\end{figure}

In Fig.~\ref{fig:mobility_and_sdh}a the gate-dependence of the 2DEG mobility is shown, measured in the same Hall bar as the data in Fig.~\ref{fig3}b of the main paper. The mobility peak is at $V_g = -2.75~\text{V}$, the same gate-voltage value at which $n_2 \approx 0$ (see Fig.~\ref{fig3}b), consistent with the interpretation of a mobility-limiting second subband being depleted. Two examples Shubnikov-de Haas oscillations of $\rho_{xx}$ is shown in Fig.~\ref{fig:mobility_and_sdh}b, at $V_g=0~\text{V}$ (the two subband regime) and $V_g = -2.5~\text{V}$ (the one subband regime).

\subsection*{Gate dependence of the superconducting properties of the 2DEG}

\begin{figure}[!ht]
\center
\includegraphics{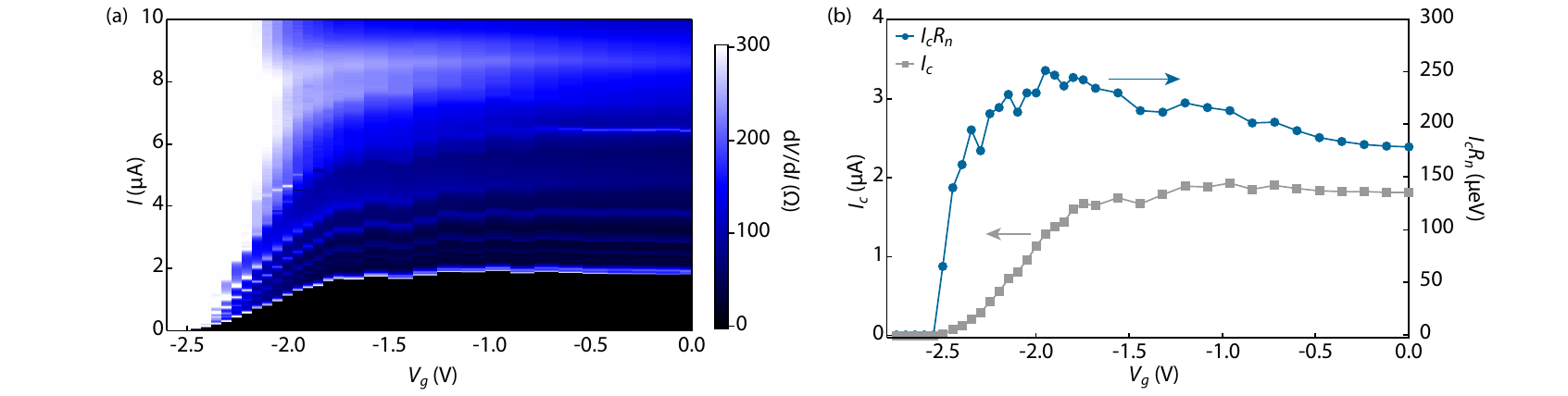}
\caption{\footnotesize{(a) Differential resistance of the junction as the current is swept, when the gate is energized. (b) The value of $I_c$ and $I_cR_n$, extracted from (a), as a function of gate voltage.}}
\label{fig:gateableIc}
\end{figure}

The gate-voltage dependence of the differential resistance of the \emph{S}-2DEG-S junction is shown in Fig.~\ref{fig:gateableIc}. At $V_g = -2.5~\text{V}$ the junction is no longer able to sustain a supercurrent, and the normal state resistance is $R_n \approx 1.7~\text{k}\Omega$. The $I_cR_n$ product is only slightly gate-voltage dependent (Fig.~\ref{fig:gateableIc}b), with a maximum $I_cR_n \sim 250~\mu\text{eV}$ at $V_g = -1.95~\text{V}$.

\subsection*{Temperature dependence of the anomalous resistance peak}

\begin{figure}[!ht]
\center
\includegraphics{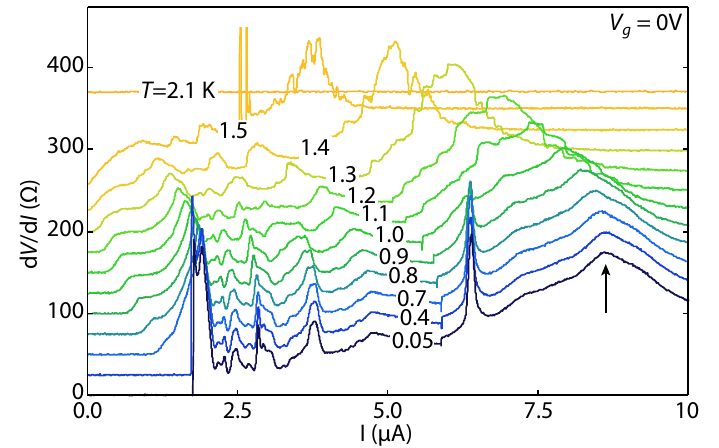}
\caption{\footnotesize{Temperature dependence of differential resistance at $V_g=0~\text{V}$. Curves successively offset by $25~\Omega$, except for $T = 0.05~\text{K}$. Vertical arrow indicates position of anomalous resistance peak.}}
\label{fig:peakVsT}
\end{figure}

Figure \ref{fig:peakVsT} shows the evolution of the differential resistance, as the temperature is increased at $V_g = 0~\text{V}$. The anomalous resistance peak in the differential resistance (highlighted with vertical black arrow) has identical temperature dependence to other superconducting features of the device. The complementary data at $V_g = -2.2~\text{V}$ is shown in Fig.~\ref{fig4} of the main paper.

\subsection*{MAR signatures in junctions with transmission close to unity}
\begin{figure}[!ht]
\center
\includegraphics{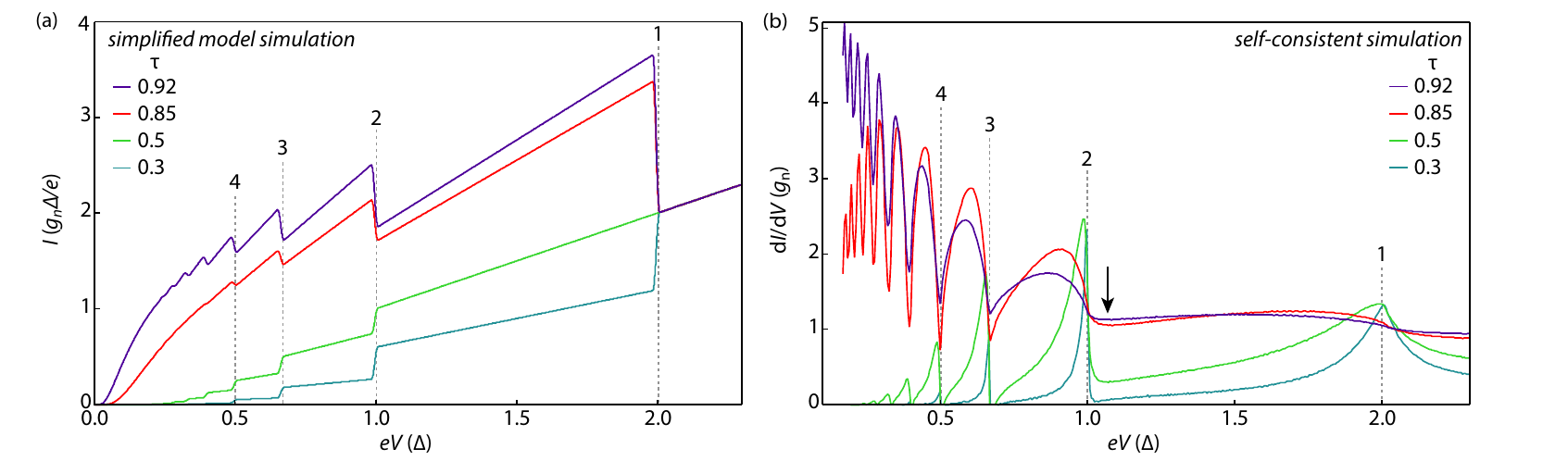}
\caption{\footnotesize{(a) Current through an \emph{SNS}--junction, from the simplified model of Eq.~\eqref{eq:approximatedcurent}, in units of $g_n \Delta /e$ ($g_n$ is the normal state conductance), for several values of transmission through the junction. (b) conductances of an \emph{SNS}-junction calculated using scattering approach for different values of the transparency $\tau$. The vertical black arrow indicates the position of the resistance peak corresponding to the experimental $p_2$ in Fig.~\ref{fig4} of the main paper.}}
\label{fig:MARtransparency}
\end{figure}

In highly transparent junctions, the $IV$ curves are qualitatively different from the opaque situation. In general, the current is a combination of the number of Andreev reflections, $n$, and the transmission $\tau$ of the junction. For the $n$'th order Andreev reflection, the particle traverses the normal region $n+1$ times, and neglecting the energy dependence of Andreev reflection probability, the current will depend on transmission as, 
\begin{equation}
I(V) \sim (n+1)\tau^{n+1} V.
\label{eq:approximatedcurent}
\end{equation}
For low $\tau$, the current thus decreases rapidly for higher order Andreev reflection processes (\emph{i.e.} increasing $n$). In contrast, for very transparent interfaces, higher order Andreev reflections will still yield an appreciable contribution to the current. This situation is demonstrated in Fig.~\ref{fig:MARtransparency}a, where we show the current in an \emph{SNS} device, calculated according to Eq. \ref{eq:approximatedcurent}. 
For low transparencies, the slope of the $I$ versus $V$ curves \emph{increases} as $n$ decreases and the current is increased at the transition from $n$ to $n-1$ Andreev reflections. As a result, the conductance of opaque junctions forms a staircase-pattern that increases in voltage with peaks at the subgap features ({\em cf.} the conductance depicted with the blue and green curves in Fig. \ref{fig:MARtransparency}b, calculated using the model of Ref.~\cite{Averin:1995do}). 
In contrast, in the transmissive junctions, the current curve exhibits an opposite pattern, which results in a declining staircase-pattern in the conductance with the peaks replaced by dips (see the purple curve in Fig.~\ref{fig:MARtransparency}b). This leads to an overall increase in the conductance {\em between} values of the voltage corresponding to integer multiples of the gap ({\em i.e.} at $V = 2\Delta/en$). 
  
In Fig.~\ref{fig4} of the main paper the peaks in differential resistance (labeled with $p_1$, $p_2$ and $p_3$) of the MAR are tracked as the temperature is increased. The energy corresponding to the second peak in the differential resistance is not identical to the value of the gap, as shown in Fig.~\ref{fig:MARtransparency}b, where the position of the resistance peak associated with the $n=2$ MAR reflection is indicated by the vertical black arrow. As the transparency is increased, the peak in resistance moves further away from integer multiples of the gap. From the simulation in Fig.~\ref{fig:MARtransparency}b the difference between the position of $p_2$ and the value of the gap is $\sim 10\%$ at $\tau = 0.92$. In the main text the value of $p_2$ is $\sim210~\mu\text{eV}$ while $\Dstar = 180~\mu\text{eV}$ in reasonable agreement with the $\sim 10\%$ difference. The correspondence between the temperature dependence of MAR features and temperature dependence of the gap is unchanged by this effect.

\subsection*{Subbands in the simulations}

In the main article we introduce the simulations used to fit the MAR features, in Fig.~\ref{fig3}a. The fit procedure takes as input a fixed number of subbands, denoted $M$, which can have a different number of modes, $N$, and transmission $\tau$. At $V_g = 0~V$ we find the optimal subband number is $M = 2$. 
\begin{figure}[!ht]
\center
\includegraphics{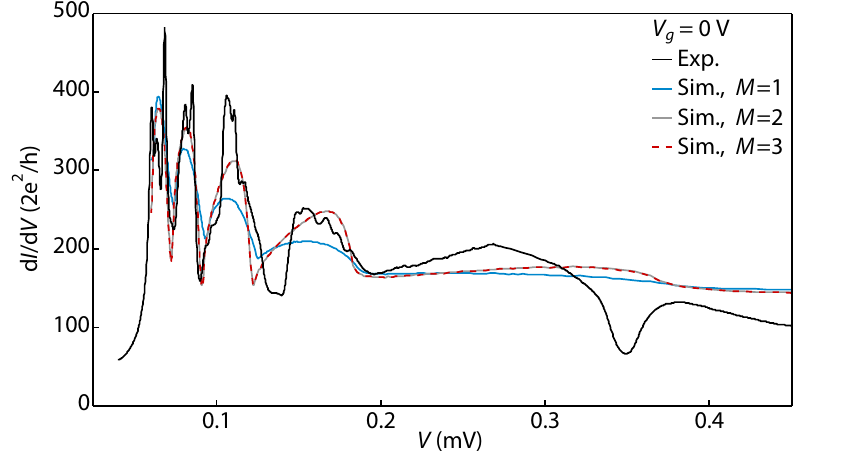}
\caption{\footnotesize{Fit to experimental data at $V_g = 0~\text{V}$ (black curve), for several values of subbands, $M$, used in the simulation.}}
\label{fig:MARatfewMs}
\end{figure}
As shown in Fig.~\ref{fig:MARatfewMs} at $M=1$ the fit is visibly worse, while for $M=3$ the optimal least-squares fit to the experimental data does not involve any modes in the third subband, i.e. $N_3 =0$. We note, that for $M = 3$ and a specific choice of the initial guess of the parameters, the fitting procedure can distribute modes among all three subbands, creating two subbands that are almost degenerate in transmission probability. When this happens the fitting errors of $N_i$ and $\tau_i$ are larger than the fitted values by several orders of magnitude, and hence we disregard such solutions. In Table~\ref{tbl:MsandNs} we list the number of modes in the optimal least-squares fit, for $M=1,2,3$.

\begin{table}[ht!]
\begin{tabular}{c|c|c}
$M=1$ & $M=2$ & $M=3$\\
\hline
$N_1 = 299,\, \tau_1=0.97$ & 
\begin{tabular}{c@{ = }l} 
$N_1$ & 199$,\, \tau_1 = 0.98$ \\ 
$N_2$ & 109$,\, \tau_2 = 0.8$ 
\end{tabular} & 
\begin{tabular}{c@{ = }l} 
$N_1$ & 185$,\, \tau_1 = 0.98$  \\ 
$N_2$ & 123$,\, \tau_2 = 0.82$  \\ 
$N_3$&0\hphantom{22}$,\, \tau_3 = \text{n.a.}$ 
\end{tabular}
\end{tabular}
\caption{\footnotesize{Optimal value of $N_i$, the number of modes in subband $i$, for different number of subbands, $M$, in the simulation.}}
\label{tbl:MsandNs}
\end{table}

\end{document}